%
%
\magnification=\magstep1
\headline={\ifnum\pageno=1\hfil\else\hfil\tenrm--\ \folio\ --\hfil\fi}
\footline={\hfil}
\hsize=6.4truein
\vsize=9.0truein
\hoffset=0.1truein
\voffset=0.1truein
\baselineskip=16pt
%
%
\tolerance=9000
\hyphenpenalty=10000
%
%
%

\font\mbf=cmmib10 \font\mbfs=cmmib10 scaled 833
\font\msybf=cmbsy10 \font\msybfs=cmbsy10 scaled 833
\font\smcap=cmcsc10
%
%
%

\textfont9=\mbf \scriptfont9=\mbfs \scriptscriptfont9=\mbfs

\textfont10=\msybf \scriptfont10=\msybfs \scriptscriptfont10=\msybfs
%
%
\mathchardef\alpha="710B
\mathchardef\beta="710C
\mathchardef\gamma="710D
\mathchardef\delta="710E
\mathchardef\epsilon="710F
\mathchardef\zeta="7110
\mathchardef\eta="7111
\mathchardef\theta="7112
\mathchardef\iota="7113
\mathchardef\kappa="7114
\mathchardef\lambda="7115
\mathchardef\mu="7116
\mathchardef\nu="7117
\mathchardef\xi="7118
\mathchardef\pi="7119
\mathchardef\rho="711A
\mathchardef\sigma="711B
\mathchardef\tau="711C
\mathchardef\upsilon="711D
\mathchardef\phi="711E
\mathchardef\chi="711F
\mathchardef\psi="7120
\mathchardef\omega="7121
\mathchardef\varepsilon="7122
\mathchardef\vartheta="7123
\mathchardef\varpi="7124
\mathchardef\varrho="7125
\mathchardef\varsigma="7126
\mathchardef\varphi="7127
\mathchardef\nabla="7272
\mathchardef\cdot="7201
%
%
\def\spose#1{\hbox to 0pt{#1\hss}}
\def\lta{\mathrel{\spose{\lower 3pt\hbox{$\mathchar"218$}}
     \raise 2.0pt\hbox{$\mathchar"13C$}}}
\def\gta{\mathrel{\spose{\lower 3pt\hbox{$\mathchar"218$}}
     \raise 2.0pt\hbox{$\mathchar"13E$}}}
%
%

%
%

%
%

%
%
%

%
%

%
%
\newcount\eqnumber
\eqnumber=1
%
\def\new{{\the\eqnumber}\global\advance\eqnumber by 1}
%
%
\def\ref#1{\advance\eqnumber by -#1 \the\eqnumber
     \advance\eqnumber by #1 }
%
%
\def\last{\advance\eqnumber by -1 {\the\eqnumber}\advance 
     \eqnumber by 1}
%
%
\def\eqnam#1{\xdef#1{\the\eqnumber}}
%
%
%
\def\refindent{\par\noindent\hangindent=3pc\hangafter=1 }

\def\aj#1#2#3{\refindent#1, AJ, #2, #3}
\def\apj#1#2#3{\refindent#1, ApJ, #2, #3}

\def\pasp#1#2#3{\refindent#1, PASP, #2, #3}

%
%
\def\refrule{\hbox to 3pc{\leaders\hrule depth-2pt height 2.4pt\hfill}}
%
%
%
\def\sect#1 {
  \bigbreak
  \centerline{\bf #1}
  \bigskip}
\def\subsec#1#2 {
  \bigbreak
  \centerline{#1.~{\it #2}}
  \bigskip}
%
%

\null

\bigskip
\bigskip

\centerline{\bf SEMI--AUTOMATIC REMOVAL OF FOREGROUND STARS}

\centerline{\bf FROM IMAGES OF GALAXIES}

\bigskip
\bigskip
\bigskip
\bigskip

\centerline{ \smcap Zsolt Frei$^1$ }

\bigskip

\centerline{ Department of Astrophysical Sciences}
\centerline{ Princeton University}
\centerline{ Peyton Hall, Princeton, NJ 08544 }

\bigskip

\centerline{ Electronic mail: frei@astro.princeton.edu }
	
\bigskip
\bigskip
\bigskip
\bigskip

\centerline{ Submitted to the Publications of the Astronomical Society 
of the Pacific }

\bigskip
\bigskip
\bigskip
\bigskip
\bigskip
\bigskip
\bigskip
\bigskip
\bigskip
\bigskip

\centerline{ \it Received 1994.............................. \ \ \ \ \ 
Accepted 199...............................  }

\bigskip
\bigskip

\leftline{ 1: Current address: Institute of Physics, E\"otv\"os University,
Budapest, Puskin u. 5-7, }
\leftline{ H-1088, Hungary }

\vfil\eject
\null\bigskip

\sect{ABSTRACT}

\bigskip

A new procedure, designed to remove foreground stars from galaxy
profiles is presented here. Although several programs exist for
stellar and faint object photometry, none of them treat star removal
from the images very carefully.  I present my attempt to develop such
a system, and briefly compare the performance of my software to one of
the well known stellar photometry packages, DAOPhot (Stetson 1987).
Major steps in my procedure are: (1) automatic construction of an
empirical 2D point spread function from well separated stars that are
situated off the galaxy; (2) automatic identification of those peaks
that are likely to be foreground stars, scaling the PSF and removing
these stars, and patching residuals (in the automatically determined
smallest possible area where residuals are truly significant); and (3)
cosmetic fix of remaining degradations in the image. The algorithm and
software presented here is significantly better for automatic removal
of foreground stars from images of galaxies than DAOPhot or similar
packages, since: (a) the most suitable stars are selected {\it
automatically} from the image for the PSF fit; (b) after star-removal
an {\it intelligent} and {\it automatic} procedure removes any
possible residuals; (c) unlimited number of images can be cleaned in
one run without any user interaction whatsoever.

\bigskip

\vfil\eject

\null\bigskip

\sect{1.~INTRODUCTION}

\bigskip

We have recently developed a catalog of digital images of 113 nearby
galaxies of various Hubble types (Frei {\it et al.}, 1996).  I used
the images in this catalog to develop techniques what enable me to
derive quantitative parameters that represent the galaxies in the
images.  These parameters will be used later to explore the sequence
of the Hubble morphological classes of galaxies, and to develop an
automated galaxy classification system (Frei, in preparation).

Foreground stars need to be carefully removed from the above images of
galaxies for a number of reasons. First, to be able to derive
quantitative ``signatures'' that truly represent the galaxy in
question, one must prevent foreground stars from contaminating the
calculations.  Second, I plan to degrade resolution in the catalog to
simulate the appearance of galaxies at higher redshifts. This will
enable me to test the limitations of the automated classification
system.  Although one could process a few hundred images
interactively, this method will not be sufficient for larger data
sets. I would like to develop an automated galaxy classification
system, and if star-removed images are needed to perform the task of
automatic classification, than an {\it automatic} star-remover should
be an integral part of that system.

Recently, several groups of researchers developed computer codes to
process digital images of stars, galaxies and other sources, all
recorded by charge-coupled devices (CCDs) or digitally scanned from
the original plates. DAOPhot developed at the Dominion Astrophysical
Observatory (Stetson 1987) is among the best known packages for
crowded-field stellar photometry.  There are also several packages
available for (faint) galaxy photometry, such as APM (Irwin and
Trimble, 1984) and FOCAS (Jarvis and Tyson, 1981).

Considering blended images is not of prime importance since images of
galaxies hopefully have no more than a handful of foreground stars,
with very little chance to find overlapping star pairs.  Due to the
low number of stars per frame, PSFs cannot be fitted with sufficient
accuracy to allow position-dependency across the field.  I can not
expect that all peaks on the image correspond to stars.  I may have
several peaks representing features in the galaxy (HII regions, spiral
arms, bright stars, globular clusters, etc). I should not remove these
as if they were in the foreground.

The first step of my method is to produce a 2-dimensional empirical
point spread function by using isolated stars in the field that are
far from the galaxy so that the backgrounds of the stars are not
affected by features in the galaxy. The most suitable stars to be used
for fitting the PSF are selected {\it automatically}.  The second step
is to identify those stars which are to be removed, both atop of the
galaxy, and near the galaxy in the frame. The software removes these
stars, and patches the residuals so that the remnants of the stars
disappear. All steps, including identification of stars to be removed,
and the intelligent patching of residuals are automatic. The last step
is the possible cosmetic fix of the bad pixels or regions (cosmic ray
events, saturation trails, etc).

In Sec. 2 I present the detailed algorithms, and in Sec. 3 I work
through a simple example to demonstrate the usual steps of processing.
Comparison of performance between DAOPhot and my method is presented
in Sec. 4. I used the {\it addstar} routine in DAOPhot to add
artificial stars to 2 different galaxies that were star-removed
previously, and removed these artificial stars with both DAOPHOT and
my program. Images of the residuals and some statistics are provided
in both cases. In Sec. 5 I outline possible future improvements.  I
will be happy to supply heavily commented C source code to those who
are interested.

\vfil\eject

\null\bigskip

\sect{2.~THE STEPS OF FOREGROUND--STAR REMOVAL}

\bigskip

My steps of star removal can be collected into three major groups,
each group containing several tasks. Some of these tasks are well
known and straightforward. I adopted and coded them for my purposes,
and I built a user interface that is most suitable for the given task.
A few others, most importantly those used to patch the residuals of a
removed star in the image, were first developed for this project. 

The steps can be summarized as follows:

\item{(1)} Fitting the PSF:
\itemitem{(1a)} Finding local peaks in the image.
\itemitem{(1b)} Selecting those peaks that are likely to be isolated stars
off the galaxy.
\itemitem{(1c)} Constructing the PSF.

\item{(2)} Removing stars:
\itemitem{(2a)} Finding local peaks that are sufficiently above the local
background.
\itemitem{(2b)} Selecting those peaks that are likely to be foreground stars.
\itemitem{(2c)} Scaling the PSF and removing stars.
\itemitem{(2d)} Examining residuals, finding the smallest possible area
where further patching is required.
\itemitem{(2e)} Patching these residuals.

\item{(3)} Cosmetic fix of images:
\itemitem{(3a)} Cosmetic fix of all degradations in the image.
\itemitem{(3b)} Removing extra stars or replacing wrongly removed stars.
\itemitem{(3c)} Saving information about the locations in the image where 
changes were made.

In this section I mention all the important parameters that must be
decided upon before this procedure will work. I will give estimated
values of these parameters which seem to be reasonable {\it a
priori}.  In the next section, where I present an actual image and go
through the steps of star removal, I will give the values of the
parameters I used for processing the given image. 

\bigskip

\centerline{2.1~{\it Fitting an Empirical Point Spread Function}}

I fit a PSF for each image separately.  Stars used for the fit are
converted to grids of the same size, and combined to form the PSF.
Due to the usually small number of stars available for my fits, I
decided to use a 2D empirical PSF for the entire image field,
independent of the location in the field.  I naturally want to find
foreground stars for the PSF fit which are far from the background
galaxy, so that the local sky around these stars is uncontaminated.
The second requirement is to find stars that are well separated from
any other objects in the image.

I determine the level of the background sky. Since I will use this sky
only to find peaks, the median value of all the pixels in the image
will serve me well. The next step is to find all isolated peaks in the
image which are above a threshold. I search all pixels to find all
local maxima, and than collect those that are separated from any other
by at least a certain preset value (usually 4-5 times the expected
FWHM of the PSF). Following this, I pick those which are above the
preset value (usually 40-50 times sigma above the sky level).
Naturally, I avoid all of those that may be high enough to be in the
nonlinear regime of the given CCD.

What remains is a set of coordinates that may represent positions of
suitable stars to be used for the PSF fitting. The following step is
to extract all regions around these peaks to grids of common size, and
to center the objects on these grids. I used parabolic centroiding
to find the true center (to subpixel precision) of the objects, and
used {\it sinc} interpolation to shift the images to be centered. I
used square grids of size several times the expected FWHM of the
PSF.

I can examine each of the candidate stars now and select those most
suitable for fitting the PSF.  First, I find the local sky around each
of the selected objects, and reject those that have local backgrounds
higher than the sky level determined earlier for the entire image,
since they may sit on a background feature, and not on a flat sky. To
find the local sky, I used a circular annulus around the object with
inner radius several times the expected FWHM of the PSF.  This
geometry ensures that any contribution to the local sky due to the
fact that the background may not be flat is canceled to first order. I
used the median value of all pixels in the annulus as local sky. My
first condition for keeping a star for the PSF fit is that the local
sky around the star in question is not significantly higher than the
background sky for the entire image. I keep all those stars which have
local sky not above the image-sky plus 75 \% of the sigma of the
image-sky. This condition is formulated as:
$$ S_{local} < S_i + 0.75 \ \sigma_i \ \ \ \ , \eqno(1) $$
where $S_i$ is the image-sky level determined earlier, $\sigma_i$ is
the standard deviation of the image-sky, and $S_{local}$ is the local
sky around an object.  Second, I calculate the FWHM for all the
objects remaining on my imaginary stack of two-dimensional grids.
After finding the median of these FWHMs, I reject objects which have
FWHMs 25 \% bigger or smaller than the median:
$$ 0.75 \ W_{median} < W_{object} < 1.25 \ W_{median} \ \ \ \ , \eqno(2) $$
where $W_{median}$ is the median of all FWHMs and $W_{object}$ is the
FWHM of the object being tested. I also order the patches according to
the FWHMs, and reject the upper quartile of the patches so ordered (if
they were not already rejected based on criteria listed earlier), in
order to use the more compact objects for the PSF fitting (this might
eliminate those stars affected more by coma):
$$ W_{object} < W_{3/4} \ \ \ \ , \eqno(3) $$
where $W_{3/4}$ is the upper quartile of all FWHMs.

Those stars still remain are going to be used for the PSF fit. I first
scale them according to the total weight of the images formed on the
grids.  I select each pixel row by row, column by column in the PSF's
grid, and identify all the pixel values corresponding to the same
pixel in each of the grids on the stack. I find the median of these
values, and place this median to the grid of the PSF. This way, each
pixel of the derived PSF is going to be the median of the
corresponding pixels on the stack.  I selected a few bright stars with
similar peak values, and supposed that all data is at least critically
sampled, consequently the median pixel is a good estimate of what I
need in the PSF.

Although all the images on the stack were individually centered, the
resulting PSF may not have the common center. I use parabolic
centroiding again to find the true center of the PSF in its grid
(usually off only by a few hundredth of the size of a pixel) and use
{\it sinc} interpolation to center the PSF on the grid.  As the last
step, the FWHM of the newly created PSF is calculated.

I designed a user interface that helps to quickly check the resulting
PSF, and modify parameters and re-fit the PSF {\it if} necessary.  The
3D surface plot of the PSF can be displayed on the screen.  The
location of the stars used for the fit can be displayed, and all the
locations of the stars once on the stack can be listed along with the
reason for rejection from the stack (conditions (1)--(3)).  Three
dimensional surface plots of any regions of the image can be
displayed, and stars can be added or removed from the stack
interactively.

\vfil\eject\null

\bigskip

\centerline{2.2~{\it Finding and Removing Stars; Patching Residuals}}

To find those stars that are to be removed, I look for peaks in the
image again. The minimum allowable separation is smaller than in the
previous case, however. This is due to two reasons: the stars found
now will not be used for the critical step of fitting the PSF, and to
be able to remove the majority of the stars automatically, I should
avoid excluding too many, just because they are not very well
separated. I found that 2-3 times the FWHM of the PSF is enough
separation for this step. 

Given the list of sufficiently separated local maxima, I select those
peaks which are above the {\it local} sky by at least a given
threshold. The local background is found in a circular annulus
centered around the peak. The inner and outer radii of this annulus is
predetermined in the units of the FWHM of the PSF. Since extended
object will contribute to the local sky in the computations (due to
the relatively small annulus) I will find higher local background than
the real value. This, in turn, will reduce the height of the peak I
determine, and I will be more likely to reject such a peak from star
removal, since the value of the peak may fall below the predetermined
threshold.

The maximum value of the peak is not restricted to be in the linear
regime of the CCD any more.  I want, however, to avoid all bleeding
columns, and therefore exclude all regions with more than a few pixels
in the saturated regime.  Extended saturated regions will be treated
separately in the final step.

To remove a given star, first I have to scale the PSF to the given
star. I extract the region where the star is and register the star on
its grid. I calculate the {\it light} corresponding to both the star
and the PSF using a special pattern of multiplicative factors
(``weights'') shown in Fig. 1.  This pattern is designed so that if
there is no light coming from the star, the total light {\it is} $0$,
{\it and} the contribution of a nonzero or nonflat local background is
canceled. The weights are positive ($+1$) in a circle around the
center of the star, and negative ($-1$) in a concentric annulus
adjacent to the central circle. If the area of the circle equals the
area of the annulus, then the pixel-by-pixel sum of this pattern is
$0$. This means that any remaining contribution from the local
background (galaxy, sky, other sources) is taken out.  Since the same
pattern of weights is used to calculate the light of both the star and
the PSF, the ratio of the two lights will be the same as the ratio of
{\it total lights} in the two images.

Comparing the two lights, I scale the PSF so that the light of the
scaled PSF matches the light of the star. I shift the center of the
scaled PSF on it's grid to match the real center of the star on the
grid of the original image, and subtract this scaled and shifted PSF
from the original image pixel-by-pixel within a predetermined radius.

After the star is removed by this method, there will be some pixels
that do not blend in smoothly with the background.  I can determine
the size of a circular region around the center of the removed star in
which the residuals are above or below a certain threshold, and I can
replace this circular region with random data bearing the same
statistical characteristics as the local background. My aim is to
replace only the smallest possible region which contains significant
residuals.

A first order plane is fitted to the data locally, in an annulus
centered around the removed star. The inner radius of this annulus is
bigger than several times the FWHM of the PSF used for removal, so
that the direction of the plane fitted will not be affected by any
errors in the star-removal process. The local sigma around the removed
star is also determined in this annulus. The region is than extracted
to a square grid, and the plane (an analytic function) is subtracted
from the data. This step ``tilts'' the area that contains the
residuals flat in the first order sense. I sum up the ``bad'' pixels
in concentric annuli around the center of the removed star on this
grid. A bad pixel is one where its value is a few times sigma above or
below $0$.  The outer and inner radii of the annuli are constantly
decreased, and the number of the bad pixels are noted. If the number
of bad pixels grow above a preset limit while decreasing the size of
the annuli, the outer radius of the annulus where this happens is
going to be the radius of the region within which the pixels has to be
replaced. 

The pixels in this circle are replaced by values drawn from a Gaussian
distribution with $0$ mean and variation which equals the variation of
the local background. The data are then ``tilted'' back to the
original slope (analytic, first order background added), and the
simulated data are replaced to the original image.

Small objects and higher order curvature in the background can
introduce more bad pixels even far from the center of the removed star
then usual. It seemed natural to set the limit for the number of
acceptable bad pixels in any given annuli {\it as the function of} the
number of bad pixels found in the outermost annulus (the one used to
determine the local sigma). I also kept a parallel condition, based on
the total number of bad pixels (in case the outer annulus is severely
affected by bad pixels, I did not want to accept all the residuals
because they are {\it relatively} better).

After some experiments, I decided to use the combination of two sets
of thresholds.  The first set of thresholds contains the following
parameters: pixels are bad if they are above or below $1$ local sigma;
the number of bad pixels (after areas of annuli carefully equalized)
should not be bigger than $2$ times the number of bad pixels in the
outer annulus, and, simultaneously, the number of bad pixels can not
be bigger than half the number of pixels in the annulus.  The second
set of thresholds are almost the same, the only difference is that the
level for a pixel being ``bad'' is increased to $2$ local sigma.

To formulate my two parallel conditions, I denote the number of bad
pixels in an annulus $N_{bp}(r)$, where $r$ is the outer radius of the
annulus (the inner radius is determined so that the area of all annuli
kept constant). I determine the number of bad pixels in the outermost
annulus ($N_{bp}(r_{outer})$), and keep decreasing the outer radii of
subsequent annuli {\it until}:
$$ N_{bp}(r) < 2 \ N_{bp}(r_{outer}) \eqno(4a) $$
and
$$ N_{bp}(r) < 0.5 \ N_p(r) \ \ \ \ , \eqno(4b) $$
where $N_p$ is the number of pixels (area) of the annulus with outer
radius $r$. If one of these conditions are not met while decreasing
$r$, this $r$ is the suggested radius of the circular region which has
to be repaired. Since I have two separate sets of conditions
(different only in the threshold used to determine the number of bad
pixels), I will obtain 2 such radii.  My experience shows that the
average of these two radii yields the best result: this average should
be the radius of the region to be repaired.

The center of a galaxy is likely to be the brightest peak in that
galaxy. If I want to be sure that I do not remove what belongs to the
galaxy, I should require that all peaks of objects to be removed
should be at least as high above the local sky as the center of the
galaxy is above the background sky determined for the entire image.
This is a very conservative condition, but this criteria is easy to
automate and removing stars with this threshold is usually enough for
subsequent processing of the galaxy-images.

If I set the threshold to be lower than outlined above, I may remove
objects that belong to the galaxy in the background. The most
important objects are HII regions, which may appear very compact and
thus resemble foreground stars. There are three criteria that can be
used to distinguish them from stars. The first is the FWHM.  For HII
regions, if they are resolved, the FWHM is {\it somewhat} larger than
for stars on the same image.  The second criteria is color. In case
images of the same galaxy are available in different passbands, the
color of objects can be examined. Since HII regions are {\it usually}
bluer than stars, the sources that are brighter on the bluer image are
{\it more likely} to be HII regions. The last criteria is position. On
images of spiral galaxies, objects close to the spiral arms are {\it
likely} to be associated with the background galaxy. It is clear that
all three conditions are probabilistic. Combining all of them may give
us results which are correct for the majority of objects. 

\bigskip

\centerline{2.3~{\it Cosmetic fix of Bad Pixels and Regions}}

The user interface I developed can be used to correct any mistakes the
above procedures may make. There is a way to remove a star (simply
pointing at it) if it was left in the image, and it is similarly easy
to put back an object if it was mistakenly removed. It is also
possible to show those stars which are about to be removed, and the
list can be modified before actually doing the removal. Parameters
described above can be adjusted if needed.

There is a utility to patch any regions of the image which may seem to
be degraded with the exact same procedure used to patch circular
regions of residuals left after star removal.  I also have a utility
to patch rectangular regions (usually saturation trails).

I store the central coordinates and radii of all the circular regions
of the image which were changed by star removal or patching.  These
data can be used for later reference, or to repeat the same steps for
a similar image (if the same source is on several images, observed in
separate time-intervals).

\vfil\eject

\null\bigskip

\sect{3.~Processing a Sample Image}

\bigskip

I selected a sample image to demonstrate the steps of processing
outlined in the previous section. This image of the galaxy M100 (NGC
4321) was obtained with the 1.5-meter reflector at the Palomar
Observatory during the night of May 4-5, 1991, in the $g$ band of the
Thuan-Gunn photometric system (Thuan and Gunn 1976), extended by Wade
{\it et al} (1979). M100 is one of the best known almost-face-on
spirals, demonstrating some of the main features of a ``grand design''
spiral, but posing still unsolved problems (Pierce 1986; Elmegreen,
Elmegreen and Seiden 1989).

The image was flat fielded, bad columns and pixels of the CCD were
repaired, and a 561 by 561 pixel region centered on the center of the
galaxy was clipped form the original CCD image of size 800 by 800
pixels. This image is shown in Fig. 2a.

The image contains 16 bit data, signed integers between $-32768$ and
$32768$. Negative numbers are not used (the data effectively contain
15 bits of information per pixel), the level of the sky was $797.1$
counts, the variation (sigma) was $29.8$ counts. I first tried to find
suitable stars for fitting the PSF automatically. The FWHM of the PSF
was estimated to be around 3 pixels, consequently the threshold for
minimum separation for those peaks to be considered in the search was
set to $10$ pixels. Minimum height of the peaks was set to sky plus
$50$ times sigma. Due to a possible nonlinearity in the CCD, the
threshold for maximum acceptable value of the peaks was set to $20000$
counts.

Thirteen peaks were found using the above thresholds. These objects
were extracted from the image and were placed on square grids. They
were centered, the local sky around them and the FWHM of each object
was computed. Five of the objects were rejected afterwards because the
local sky was too high (see condition (1)). Five other objects were
rejected because of problems with the FWHMs (see conditions (2) and
(3)). The remaining 3 stars were used to fit the PSF. I decided to use
more stars and re-fit the PSF.  After lowering the minimum height for
an acceptable star to sky plus 40 sigma, 6 suitable stars were found
for the PSF fit.  These stars are marked in Fig. 2b.

The following step was the actual star-removal using the PSF just
constructed. For the purpose of this demonstration I decided to set
the lower limit of the peaks to be considered for star removal below
the peak of the galaxy (which was {\it very} high at $18225$ counts),
at $2500$ counts.  There were 25 peaks on the image higher than this
value, but only 17 of them were actually $2500$ counts higher than the
{\it local} sky around the corresponding peak. These 17 stars are
marked on Fig. 2c.  The square bracket close to the edges of the image
in Fig. 2c encloses the region that is used to look for peaks.
Anything outside this region is too close to the edges of the image to
be able to fit the PSF.  After removal and patching the residuals, no
visible traces of any of the stars were left on the image, which is
shown in Fig. 2d.  The four peaks closest to the center of M100 are
most likely to be HII regions, and not foreground stars since they are
much brighter in the $i$ band image, than in the $g$ band image (I
compared this image to the H$\alpha$ image of M100 in Hodge and
Kennicutt (1983), and the four peaks in question appear to be HII
regions, indeed).  However, the PSF are as narrow as the PSFs of stars
on this image, so I decided to remove them to show how my program
performs on stars ``sitting'' on features of the background galaxy.

Although it is not demonstrated here, the lower limit for the height
of the stars to be removed can be set even lower, and additional
sources can be removed individually using the user interface developed
for this purpose. 

\vfil\eject

\null\bigskip

\sect{4.~Performance Compared with DAOPhot}

\bigskip

DAOPhot is one of the best known photometry packages available for
crowded stellar field photometry.  I intend to demonstrate that
removing foreground stars from images of background galaxies is a {\it
different} task from processing crowded stellar fields, and
consequently, it was necessary to develop a system tailored for my
needs.

I chose two images of spiral galaxies, the nearly edge on NGC 4192 and
the nearly face on NGC 4535. Both of these galaxies are part of our
digital galaxy catalog, and were observed with the 1.5-meter reflector
at the Palomar Observatory on the night of May 4-5, 1991.  I used the
$i$ passband of the Thuan-Gunn photometric system for these images.

I completely cleaned the images of any unresolved sources, then added
artificial stars to both of them using DAOPhot, and removed those
stars using both DAOPhot and my software. After star removal, I
subtracted the original, clean image from the star-removed images, and
statistically compared the remaining residuals.
 
For creating artificial stars, I used the original (not star-removed)
images of the two galaxies obtained at Palomar to fit PSFs with
DAOPhot. This PSF, determined on the original image of the given
galaxy was stored in a file, and this PSF was used to add artificial
stars to the corresponding clean image (cleaning was done with my
software between these two steps). This way the characteristics of the
stars added to the image is the same as the characteristics of the
galaxy in the background. I set up a very simple grid of 16 stars, all
of the same magnitudes, and used DAOPhot's {\it addstar} to add stars
to the images. The combined images are shown in Figs. 3a and 4a for
NGC 4192 and NGC 4535, respectively.

To remove stars with DAOPhot, I used the newly created images to fit a
new PSF, based on the stars now on the image. If I let DAOPhot use all
the bright stars for fitting the PSF, it will blindly select some of
those sitting atop of the galaxy for this fit. I removed stars with a
PSF fitted this way with DAOPhot, then used my program to identify
those stars which are off the galaxy, fed this information back to
DAOPhot, and fitted a new PSF using the ``suitable'' stars only.  Star
removal with this PSF is much more successful.  I have to stress that
my program - unlike DAOPhot - can automatically find those stars which
are off the galaxy and which are consequently most suitable for
fitting the PSF. I used my program to {\it help} DAOPhot, so that I
can compare the star removal part of the two softwares.

After star removal with DAOPhot, I subtracted the original, clean
images. The residuals are shown in Figs. 3b and 4b for the simple PSF
fit within DAOPhot, Figs. 3c and 4c show residuals with the better
PSF, where I used my software to determine suitable stars for fitting
the PSF.  Every pixel that is not affected by the star-removal
procedure is $0$ in the subtracted images, and they are shown in white
in the figures. To be able to show differences above {\it and} below
$0$ on these images, I plotted the absolute value of each pixels.

Please note, that the stretch of grey scales are different for the
images obtained with the different methods. On all figures, white
represents $0$ counts, but the grey scale is stretched to different
maximum values (pixels with the maximum count, and all pixels
containing a value above it, are shown in black).  The stretch is to
$500$ and $300$ maximum counts on Figs. 3b and 4b, respectively, while
only to $30$ in both Figs. 3c and 4c.

To quantify the results, I calculated the ``power'' left in the
residuals. To do this, I simply summed the squared pixel values for
every pixel in the residual images, divided by the number of pixels,
and took the square-root of the result:
$$ P =\sqrt{ {1 \over N} \sum_{i}^N p_i^2 } \ \ \ \ ,
\eqno(5) $$
where $P$ is the residual ``power'' per pixel, $N$ is the number of
pixels, and $p_{i}$ is a pixel-value.

In Table 1 I show the power per pixel after star removal with the
simple PSF ($P_1$) and with better PSF ($P_2$) for both images.  It is
obvious, that if I do not add and remove stars, just simply subtract
two featureless regions of an image with similar sky values and
variations, I will detect ``power'' due to the noise in the image
using the method described by (5).  If the variation of the sky is
$\sigma$, then due to the propagation of errors, the expected
``power'' per pixel is $\sqrt{2}\sigma$. This expected power per pixel
due to the variation in the sky is also shown in Table 1.

With presenting these numbers I intend to show the importance of
finding suitable stars for the PSF fit - what is done by my program -
if there is a relatively faint but extended object in the
background. It is interesting to note that the residual power can be
less than the expected power due to the variation of the sky.  This is
due to the fact that I subtracted the very same original image to
which artificial star were added. In theory, if the PSF used for
adding stars and the PSF used later for star removal are perfectly
matched, with this method we can get $0$ residual power.

I used my method to remove the artificial stars from the same images,
and subtracted the original, clean image after star-removal.  To be
able to compare the star removal by PSF fitting (without patching the
residuals) between DAOPhot and my method, I first switched off
patching in my software. The residuals are shown in Figs. 3d and 4d,
and the residual powers ($P_3$) are in Table 1. I also used patching,
the corresponding residuals are in Figs. 3e and 4e, the residual
powers are $P_4$ in Table 1.

It is clear from these numbers ($P_2$ vs. $P_3$) that my star removal
method (without patching) gives comparable residual powers than
DAOPhot, when the PSF is fitted using the same set of stars. That is,
the star removal steps are similar in both programs. However, the
overall performance is much better for my code since it automatically
finds suitable stars for the PSF fit, and after removal, it uses a
special algorithm to patch possible residuals.  The difference between
$P_1$ and $P_2$ clearly suggests that finding suitable stars for the
PSF fit is very crucial.

The residual power after patching is substantially larger than without
patching. These numbers are misleading, since with patching I introduce
random noise to the data. Even if no residuals of the removed stars
are left in the image, I expect some residual power due to the random
variation in the sky (see the last line in Table 1).  Patching
certainly removes residuals of subtracted stars, and produces an image
where traces of removed foreground stars are hardly noticeable (see
Fig. 2d).
 
\vfil\eject

\null\bigskip

\sect{5.~Conclusions and Possible Improvements}

\bigskip

The small residuals that remained in the images after star removal
presented in the previous section demonstrate that my methods can
clean foreground stars from images of galaxies to a satisfactory
degree. The level of automation unfortunately is not very high at this
stage. I can run the procedure automatically and achieve results
better than with DAOPhot, but dim stars are left in the image which
can be removed only interactively. Systematic studies of statistical
properties of these objects are required to further automate my
procedure.

The star-removal procedure itself can be further improved. The PSF
fit can be modified to work for undersampled images. If a great
number of stars are available for the PSF fit, roundness of these
stars can be introduced as a new criteria to select the best stars for
the fit. The procedure can be expanded to handle blended images of
groups of nearby stars. A special procedure could be introduced to
treat bright stars in the nonlinear regime of the CCD better.  FWHMs,
colors, and positions of objects which might belong to the background
galaxy should be evaluated automatically to better guess whether they
are background objects or foreground stars.

I especially thank James Gunn for his continuous support and good
suggestions throughout this project. I acknowledge useful discussions
with Puragra (Raja) Guhathakurta and Neil Tyson. I also thank the
anonymous referee for making this paper much more readable with many
useful suggestions. This research was supported in part by NSF through
grant no. AST-9100121, and by OTKA through grant no. F 17150. Z. F. is
a {\it Magyary Fellow}.

\vfil\eject

\null\bigskip

\sect{REFERENCES}

\apj{Elmegreen, B. G., Elmegreen, D. M., \& Seiden, P. E. 1989}{343}{602}

\aj{Frei, Z., Guhathakurta, P., Gunn, J. E., \& Tyson, J. A. 1996}{111}{174}

\aj{Hodge, P. W., \& Kennicutt, R. C. 1983}{88}{296}

\aj{Irwin, M. J., \& Trimble, V. 1984}{89}{83}

\aj{Jarvis, J. F., \& Tyson, J. A. 1981}{86}{476}

\aj{Pierce, M. J. 1986}{92}{285}

\pasp{Stetson, P. B. 1987}{99}{191}

\pasp{Thuan, T. X., \& Gunn, J. E. 1976}{88}{543}

\pasp{Wade, R. A., Hoessel, J. G., Elias, J. H., \& Huchra, J. P. 1979}{91}
{35}

\vfil\eject

\null\bigskip

\sect{FIGURE CAPTIONS}

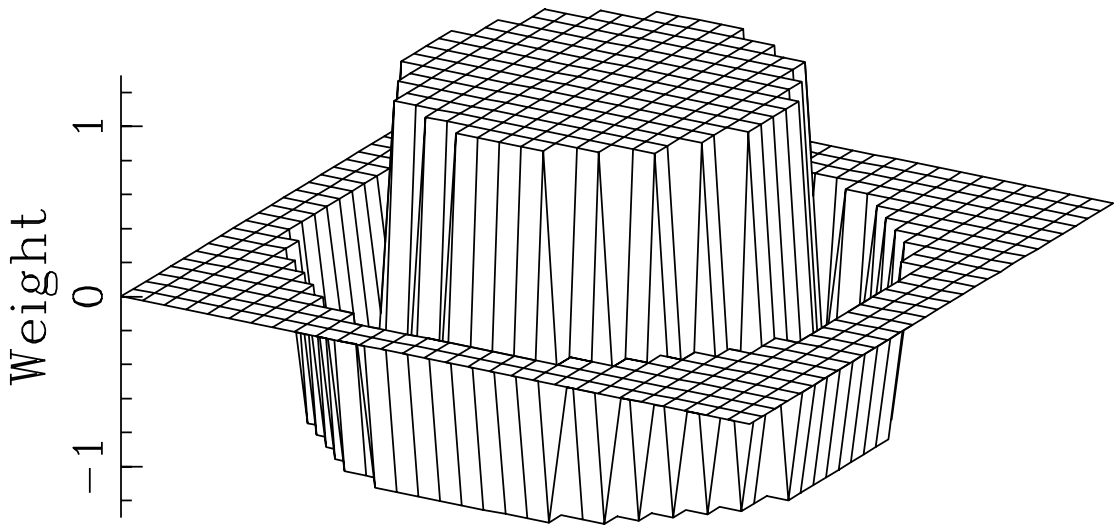
{\bf Figure 1.} Suface plot of the template of weights used to
calculate the light corresponding to stars and to the PSF. Due to
symmetry, any nonzero or nonflat sky cancels (the total ``weight'' of
this pattern is $0$).

\medskip

{\bf Figure 2a.} Image of M100 (NGC 4321) in the Thuan-Gunn $g$ band.
The background galaxy and all foreground stars before star removal are
shown.

\medskip

{\bf Figure 2b.} The same galaxy as in Fig. 2a. 6 stars to be used for
fitting the PSF are marked.

\medskip

{\bf Figure 2c.} The same galaxy as in Fig. 2a.  17 stars to be
removed are marked, still before star removal.

\medskip

{\bf Figure 2d.} The image of M100 after the 17 stars marked in Fig.
2c were removed and the residuals were patched.

\medskip

{\bf Figure 3a.} The image of galaxy NGC 4192 in the Thuan-Gunn $i$
band, with 16 bright artificial stars superimposed. The original image
was cleaned of stars first, and 16 artificial stars were added later.
The square bracket marks the region that is clipped and enlarged for
Figs. 3b-3e.

\medskip

{\bf Figure 3b.} Image of residuals after star removal with DAOPHOT,
using all bright stars for the PSF fit.  The 16 artificial stars shown
on Fig. 3a were removed and the clean image was subtracted. The
absolute value of pixels are displayed. The grey scale is stretched
between 0 (white) and 500 (black). The square bracket in Fig. 3a marks
the region that is shown here.

\medskip

{\bf Figure 3c.} Image of residuals after star removal with DAOPHOT,
using well selected stars for the PSF fit. The 16 artificial stars
shown on Fig. 3a were removed and the clean image was subtracted.  The
absolute value of pixels are displayed. The grey scale is stretched
between 0 (white) and 30 (black). Please note, that the dynamic range
of the residuals is much smaller than in Fig. 3b. The square bracket
in Fig. 3a marks the region that is shown here.

\medskip

{\bf Figure 3d.} Residuals after star removal using our star
subtraction (no patching). The 16 artificial stars shown on Fig. 3a
were removed and the clean image was subtracted. The absolute value of
pixels are displayed. The grey scale is stretched between 0 (white)
and 50 (black). The square bracket in Fig. 3a marks the region which
is shown here.

\medskip
 
{\bf Figure 3e.} Residuals after star removal using our method (with
patching). The 16 artificial stars shown on Fig. 3a were removed and
the clean image was subtracted. The absolute value of pixels are
displayed. The grey scale is stretched between 0 (white) and 100
(black). The square bracket in Fig. 3a marks the region which is shown
here.

\medskip

{\bf Figure 4a.} The image of galaxy NGC 4535 in the Thuan-Gunn $i$
band, with 16 bright artificial stars superimposed. The original image
was cleaned of stars first, and 16 artificial stars were added later.
The square bracket marks the region that is clipped and enlarged for
Figs. 4b-4e.

\medskip

{\bf Figure 4b.} Image of residuals after star removal with DAOPHOT,
using all bright stars for the PSF fit. The 16 artificial stars shown
on Fig. 4a were removed and the clean image was subtracted. The
absolute value of pixels are displayed. The grey scale is stretched
between 0 (white) and 300 (black). The square bracket in Fig. 4a marks
the region that is shown here.

\medskip

{\bf Figure 4c.} Image of residuals after star removal with DAOPHOT,
using well selected stars for the PSF fit. The 16 artificial stars
shown on Fig. 4a were removed and the clean image was subtracted.  The
absolute value of pixels are displayed. The grey scale is stretched
between 0 (white) and 30 (black). Please note, that the dynamic range
of the residuals is much smaller than in Fig. 4b. The square bracket
in Fig. 4a marks the region that is shown here.

\medskip

{\bf Figure 4d.} Residuals after star removal using our star
subtraction (no patching). The 16 artificial stars shown on Fig. 4a
were removed and the clean image was subtracted. The absolute value of
pixels are displayed. The grey scale is stretched between 0 (white)
and 50 (black). The square bracket in Fig. 4a marks the region which
is shown here.

\medskip
 
{\bf Figure 4e.} Residuals after star removal using our method (with
patching). The 16 artificial stars shown on Fig. 4a were removed and
the clean image was subtracted. The absolute value of pixels are
displayed. The grey scale is stretched between 0 (white) and 100
(black). The square bracket in Fig. 4a marks the region which is shown
here.

\medskip

\bye